\documentclass[a4paper, 11pt]{article}
\pdfoutput=1

\usepackage{url}
\usepackage{natbib}
\usepackage{mathtools}
\usepackage[margin=2.54cm]{geometry}

\renewcommand{\d}{\mathrm{d}}

\title{A Re-examination of Ellipticity Corrections for Seismic Phases}

\author
    {Stuart Russell\textsuperscript{1}, John F. Rudge\textsuperscript{1}, Jessica C. E. Irving\textsuperscript{2}, Sanne Cottaar\textsuperscript{1} \\ \small{1. University of Cambridge. Email: sr895@cam.ac.uk} \\ \small{ 2. University of Bristol}}
    
\date{}

\begin{document}

\maketitle

\section*{Summary}

The Earth’s ellipticity of figure has an effect on the travel times of seismic waves over teleseismic distances. Tables of ellipticity corrections and coefficients have been used by seismologists for several decades, however due to the increasing variety and complexity of seismic phases in use, current tables of ellipticity coefficients are now outmoded and incomplete. We present a Python package, EllipticiPy, for the calculation of ellipticity corrections that removes the dependence on pre-calculated coefficients at discrete source depths and epicentral distances. EllipticiPy also facilitates the calculation of ellipticity corrections on other planetary bodies. When applied to both Earth and Mars, the magnitudes of ellipticity corrections are on the order of single seconds and are significant for some seismic studies on Earth but remain negligible on Mars due to other greater sources of uncertainty.

\subsection*{Keywords}
    Body Waves; Computational Seismology; Theoretical Seismology; Planetary Seismology

\section{Introduction}

It has been known for many decades that the ellipticity of figure of the Earth has a significant effect on the travel times of seismic waves propagating over teleseismic distances \citep[e.g.][]{Jeffreys1935, Bullen1937}. \citet{Bullen1937} tabulated ellipticity corrections for P and S waves for combinations of source co-latitude, azimuth and epicentral distance to the receiver. \citet{Dziewonski1976} demonstrated the additional importance of source depth and substantially advanced the mathematics underpinning the calculation of ellipticity corrections by representing the correction as a degree 2 spherical harmonic expansion with calculable coefficients that depend only on phase, source depth and epicentral distance. \citet{Dziewonski1976} also presented tables of these coefficients for commonly used seismic phases in the radially symmetric Parametric Earth Model \citep[PEM, ][]{Dziewonski1975}.

As the number and complexity of seismic phases in use continued to increase, a re-examination of ellipticity corrections by \citet{Kennett1996} advanced the mathematics to allow the calculation of coefficients for diffracted phases. \citet{Kennett1996} used the reformulation of \citet{Doornbos1988} coupled with the tau-spline procedure of \citet{Buland1983} to produce tabulated coefficients for the most comprehensive list of coefficients yet. Furthermore  \citet{Kennett1996} produced a freely available Fortran package (including \texttt{ellip} and \texttt{ttimel}) to allow seismologists to calculate ellipticity corrections based on interpolating tables of pre-calculated coefficients. Additionally, \citet{Kennett1996} demonstrated how coefficients for more complex phases can be calculated by a weighted sum of those for existing phases. In principle this allowed the calculation of ellipticity corrections for any seismic phase, however has the disadvantage of interpolating coefficients calculated at discrete distances and source depths for a single model.

Tables of travel times for individual velocity models have long been obsolete in many applications and have been replaced with software that allows the calculation of ray-theoretical travel times for a given ray path. One of the foremost software for this is the TauP Toolkit \citep{Crotwell1999} which has since been incorporated into ObsPy \citep{Beyreuther2010}, giving maximum utility to the modern seismologist.
Nevertheless, interpolating tables is more efficient than ray path integrals and therefore tables are still used where efficiency is the priority, for example when handling large numbers of source-receiver pairs.

In this study, we present a summary of the theory of calculating ellipticity corrections for seismic phases and identify discrepancies in the tabulated coefficients of \citet{Kennett1996}. Subsequently we present a software package, EllipticiPy, for the calculation of ellipticity corrections that is designed to work alongside ObsPy TauP, allowing the calculation of corrections for any ray path in any velocity model, including of other planets. We then present applications of this package to both Earth and Mars.

\section{Theory}

Given a planet's ellipticity of figure, $\epsilon(r)$, the relative perturbation of a surface of constant density is

\begin{equation} \label{ellipticity}
    r^{-1}\delta r (\vartheta, \phi) =  \epsilon(r)\left(1/3 - \cos^{2}\vartheta\right) \\ = -\frac{2}{3}\epsilon(r) P_{2}^{0}(\cos\;\vartheta)
\end{equation}

\noindent where $r$ is radius, $\delta r$ is the perturbation from $r$ due to ellipticity,  $P_{2}^{0}$ is the associated Legendre polynomial of degree 2 and order 0, $\vartheta$ is co-latitude and $\phi$ is longitude.

The time correction to be added to a seismic travel time prediction for a 1D spherical model to account for the planet's ellipticity of figure can similarly be represented as a degree 2 spherical harmonic \citep{Dziewonski1976},

\begin{equation} \label{tau_legendre_addition}
    \delta t = \sum_{m=0}^{2} \sigma_{m} P_{2, m}(\cos \vartheta_{0}) \cos m \zeta
\end{equation}

\noindent where $\vartheta_{0}$ is the source co-latitude, $\zeta$ is the azimuth from source to receiver and $P_{2, m}$ are Schmidt semi-normalised associated Legendre polynomials of degree 2 and order $m$. $\sigma_{m}$ are calculable coefficients that are dependent only on phase, distance and source depth,

\begin{equation} \label{tau}
        \sigma_{m} = \sum_{i} \left\{ {\int_{q_{0}}^{q_{1}} \left( \xi - 1 \right) \epsilon \lambda_m (\theta) \ \d q} \right\}_{i} \\ -  \sum_{j} \left\{ \epsilon \lambda _{m} (\theta) \left[q\right]_-^+ \right\}_{j} \\ \mp \sum_{k} \left\{ \epsilon \lambda _{m} (\theta) q \right\}_{k}
\end{equation}

\noindent where $i$ is a sum over the continuous regions of the ray path and $j$ and $k$ are sums over discontinuities where the ray is transmitted and reflected, respectively. In the sum over $k$, the preceding minus/plus refers to top-side/bottom-side reflections, respectively. $\eta$ and $q$ are ray slowness and vertical slowness, respectively

\begin{gather}
    \eta = \frac{r}{v}, \\ q = \sqrt{\eta^{2} - p^{2}},
\end{gather}

\noindent where $r$ is radius, $v$ is velocity and $p$ is the ray parameter and $\xi$ is a convenient parameter from \citet{Bullen1963}

\begin{equation} \label{xi}
    \xi = \frac{\eta}{r} \left( \frac{\d \eta}{\d r} \right)^{-1} = \frac{\d \ln r}{\d \ln \eta}.
\end{equation}

The sub-scripted vertical slownesses, $q_{0}$ and $q_{1}$ are the vertical slownesses at the start and end of the continuous ray segments. $\lambda_{m}$ is a distance dependent variable containing a Schmidt semi-normalised associated Legendre polynomial of degree 2 and order $m$

\begin{equation} \label{lambda}
    \lambda_{m} (\theta) = - \frac{2}{3} P_{2, m}(\cos \theta).
\end{equation}

\begin{figure*}
    \centering
    \includegraphics[width=15cm]{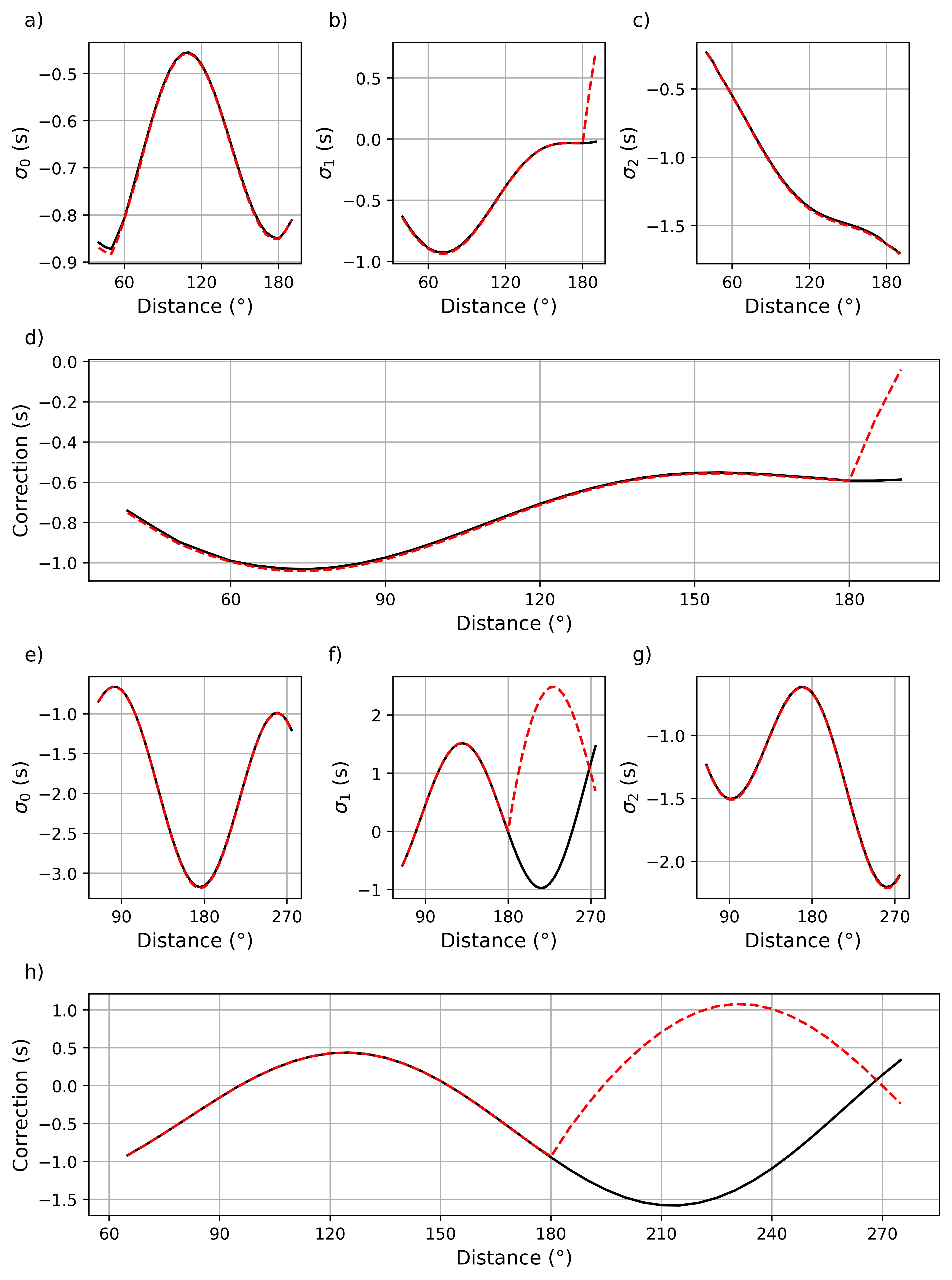}
    \caption{a), b) and c) show the values of $\sigma_{0}$, $\sigma_{1}$, $\sigma_{2}$, respectively, as a function of epicentral distance for a PP wave with a source depth of 200 km. The coefficients and corrections of \citet{Kennett1996} are shown in dashed red and those from this publication are in solid black. d) shows the corrections that these coefficients give as a function of epicentral distance for a PP wave with a source depth of 200 km, source latitude of 45\textdegree\ and an azimuth of 30\textdegree. e), f), g) and h) show the same but for an SKKS\textsubscript{ac} wave with the same source parameters.}
    \label{incorrect}
\end{figure*}

\begin{figure*}
    \centering
    \includegraphics[width=15cm]{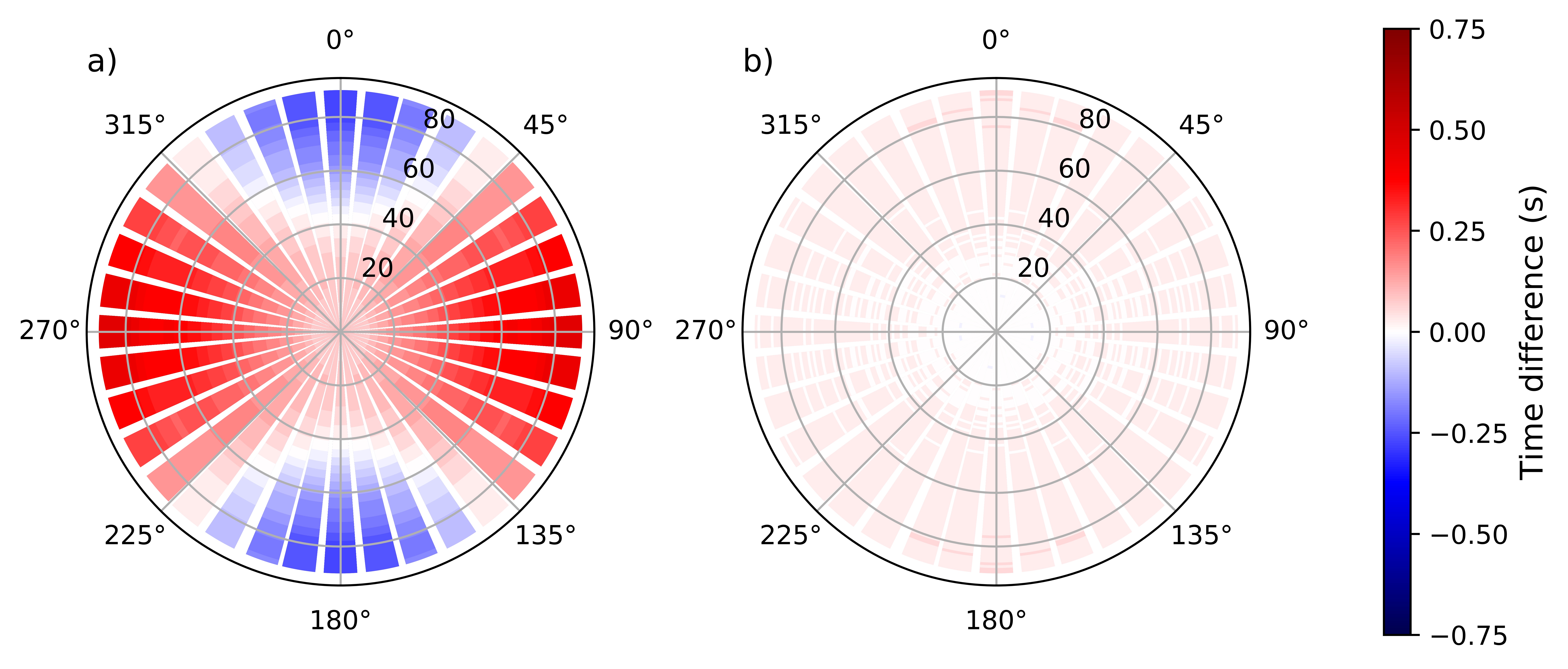}
    \caption{Rose diagrams showing the magnitude of the time difference between the elliptical and spherical SPECFEM3D synthetics for a). uncorrected and b). corrected P waveforms. The angle from the centre point is azimuth from the source and the radius is epicentral distance from the source. The residual once corrected is no more than 0.05 s at all distances and azimuths.}
    \label{SPECFEM}
\end{figure*}

\citet{Dziewonski1976} used an integral over distance in equation \eqref{tau}, however this formulation suffers from a discontinuity in distance at the centre of the Earth. Equation \eqref{tau} instead uses the formulation of \citet{Doornbos1988}, which integrates over vertical slowness without sacrificing the convenience of the original formulation. A full derivation of these equations is given in Section 1 of the supplementary materials.

Ellipticity coefficients for most major phases already exist in published tables \citep{Kennett1996} for the ak135 model \citep{Kennett1995}, however some of these coefficients are not correct. Figure \ref{incorrect} compares the coefficients and corrections for PP and SKKS\textsubscript{ac} waves from this publication and those of \citet{Kennett1996}. Below 180\textdegree\ distance there is very good agreement between our study and theirs, however beyond 180\textdegree\ they diverge. This divergence is due to a discontinuity in the gradient of $\sigma_{1}$ in the coefficients of \citet{Kennett1996}. This discontinuity has a major effect on the value of the ellipticity correction and is incorrect (Kennett, personal communication). It is only for up-going p and s waves and at distances greater than 180\textdegree\ for other phases that the $\sigma_{1}$ coefficient of \citet{Kennett1996} is incorrect. It is therefore anticipated that the vast majority of publications will not be affected by these errors as most seismological publications are concerned with minor arc phases and few studies correct up-going p and s.

To benchmark our corrections, SPECFEM3D \citep{Komatitsch2002I,Komatitsch2002II} synthetics were created in the case of a spherical and an elliptical mesh with a minimum period of approximately 7 seconds. The time difference between waveforms at the same azimuth and epicentral distance was measured by cross-correlating the windowed waveforms of a particular phase. Figure \ref{SPECFEM} shows the measured time differences from uncorrected and corrected synthetic waveforms for a direct P wave. Corrections are applied by adding the correction calculated from equation \eqref{tau_legendre_addition} to the arrival time of the spherical synthetic waveforms. For direct P waves, the corrections of this publication are equal to those of \citet{Kennett1996}. When corrected, the residuals of the corrected waveforms are extremely close to zero and have no dependence on azimuth or distance. 

\section{EllipticiPy: a Python package for the calculation of ellipticity corrections}

In light of the incorrect coefficients in the existing Fortran software package of \citet{Kennett1996}, and that the phases used in seismology studies are advancing beyond the scope of current tables, we have created a Python package, EllipticiPy, that calculates an ellipticity correction for a given ray path in any given velocity model by application of the trapezoidal rule to equation \eqref{tau}. EllipticiPy calculates the values of $\epsilon$ from the density profile in the given one-dimensional model; a derivation of how $\epsilon$ is calculated can be found in Section 2 of the supplementary materials.

EllipticiPy is designed as a companion to ObsPy TauP which allows the calculation of a spherical travel time for any ray path in any velocity model. The elliptical travel time is equal to the sum of the spherical Earth travel time from ObsPy TauP and the ellipticity correction from EllipticiPy. While EllipticiPy is designed to be used alongside ObsPy, in certain applications where large numbers of source-receiver pairs are used, interpolating tables of pre-calculated coefficients may be more efficient and as such, a function to produce tables of coefficients for specified phases is available in the package. Moreover, tables of coefficients in the exact same form as \cite{Kennett1996} have been computed (see Section 4 of the supplementary materials). Another package based on dynamic ray tracing, \texttt{raydyntrace} \citep{tian2007}, can, amongst other routines, calculate ellipticity corrections, however it is not as versatile for that specific purpose as EllipticiPy. The use of ObsPy TauP within EllipticiPy enables complex phases to be handled easily, including diffracted phases, and EllipticiPy can be applied to other planets.

\begin{figure*}
    \centering
    \includegraphics[width=14cm]{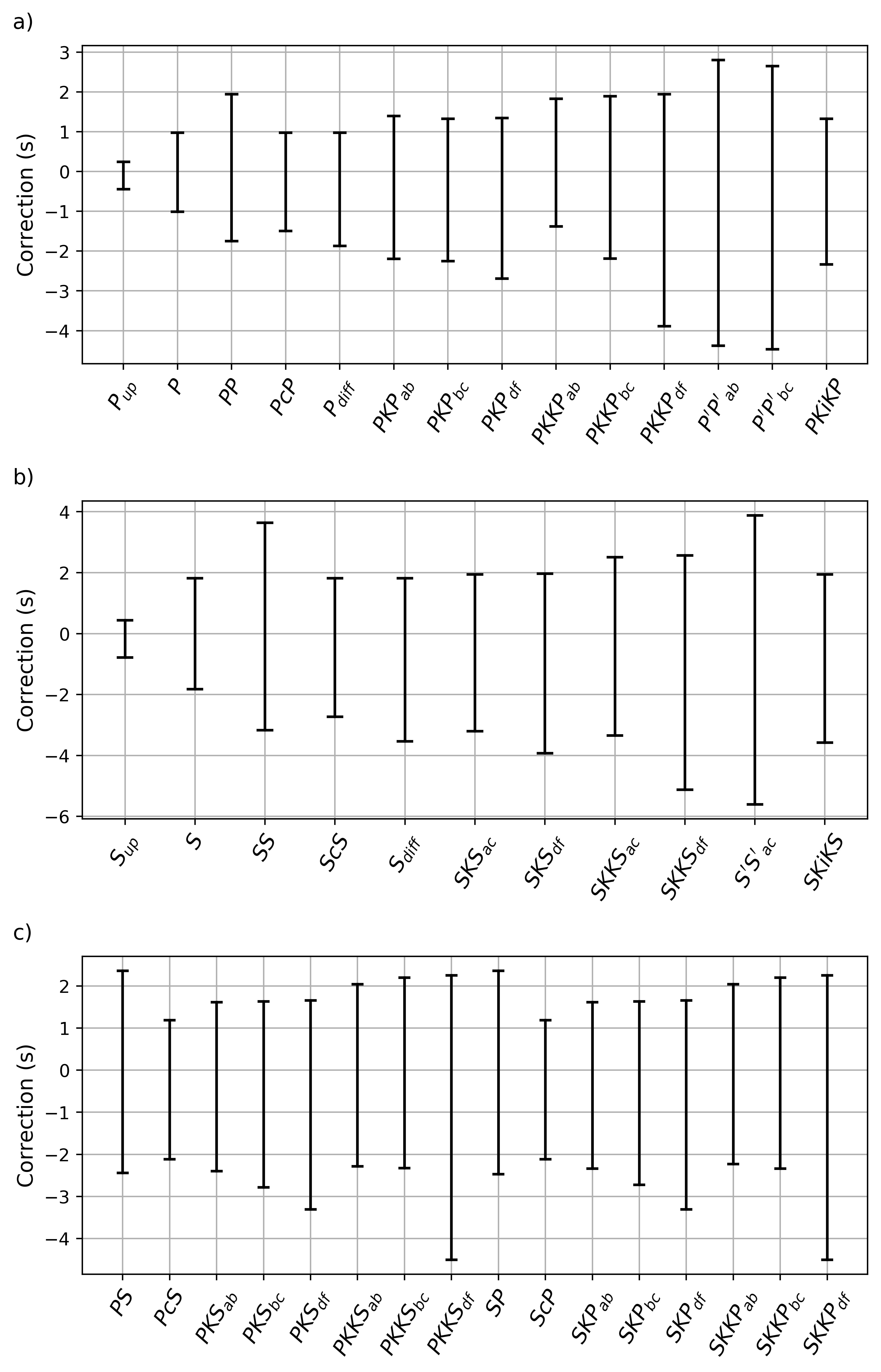}
    \caption{Minimum and maximum ellipticity corrections in seconds for commonly used seismic phases. P phases are shown in a), S phases in b) and converted phases in c). Tabulated values can be found in Section 3 of the supplementary materials. Corrections for depth phases can also be calculated but are not shown here.}
    \label{Earth corrections}
\end{figure*}

\begin{figure*}
    \centering
    \includegraphics[width=14cm]{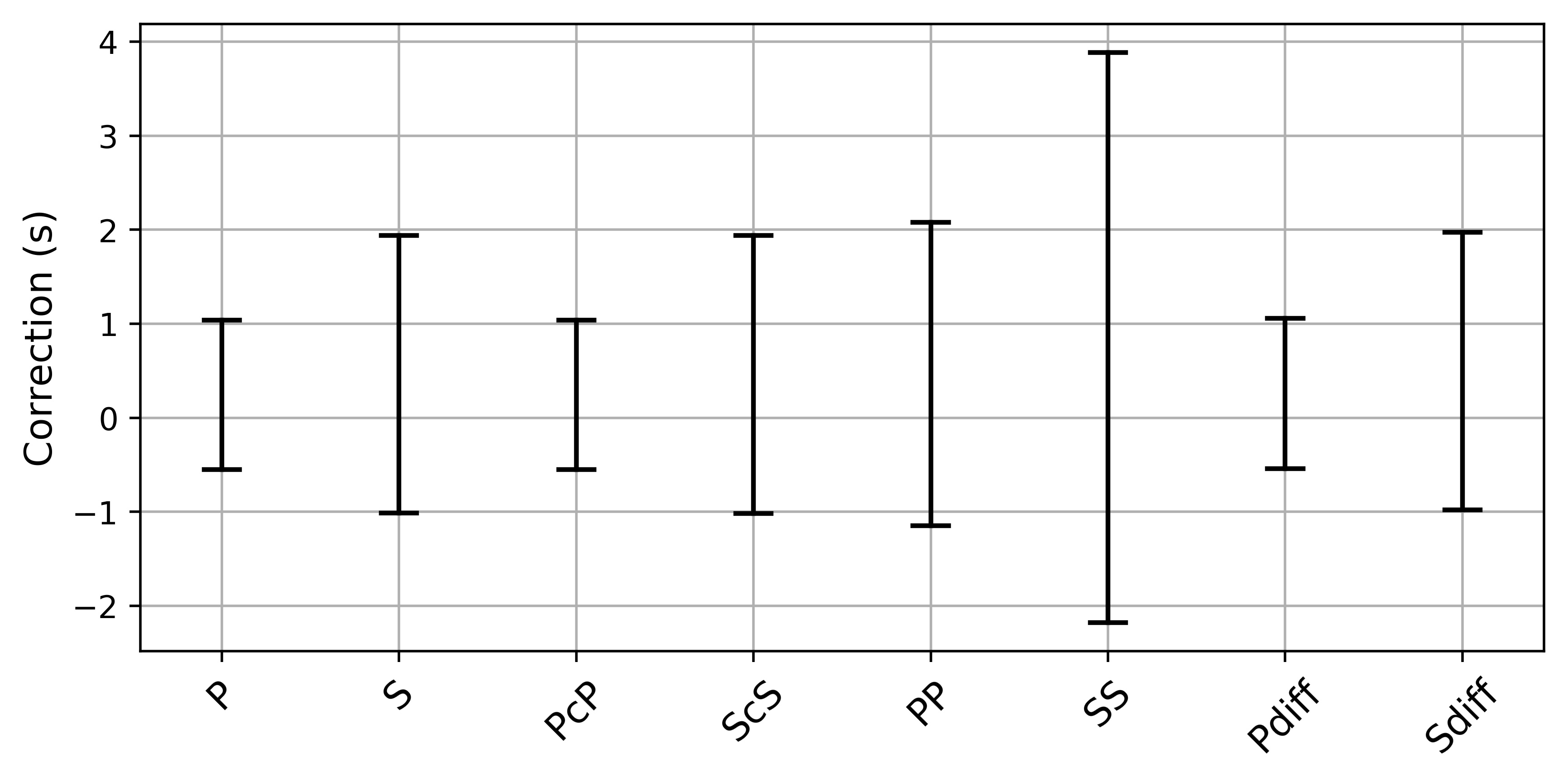}
    \caption{Minimum and maximum ellipticity corrections in seconds for phases potentially detectable on Mars at the InSight Lander location (4.50\textdegree N, 135.62\textdegree E).}
    \label{Mars corrections}
\end{figure*}

In recent decades, seismology has been applied to other bodies in the solar system and, similar to ObsPy TauP, EllipticiPy can also be applied to velocity models of other planets. Seismic waves have now been used to study the internal structure of the Moon \citep[e.g.][]{Latham1973, Garcia2019} and more recently Mars \citep[e.g.][]{Knapmeyer2021, Khan2021, Stahler2021}. These bodies have a different ellipticity of figure to the Earth which EllipticiPy calculates in the same way as for Earth (see Section 2 of the supplementary materials). As extra-terrestrial seismology advances it is likely that elliptical effects will need to be considered in the future. It should be noted that the Darwin-Radau equation \citep{Bullen1975} must hold true in order for the accurate calculation of $\epsilon$, requiring that the body is in hydrostatic equilibrium and is an ellipsoid of revolution.

\section{Applications}

\subsection{Ellipticity corrections on Earth}

Ellipticity corrections on Earth are on the order of single seconds. Figure \ref{Earth corrections} presents the minimum and maximum ellipticity corrections for common seismic phases. These can be found in tabular format in Section 3 of the supplementary materials. These corrections have been calculated by a systematic search over source depths from 0 km to 700 km, source latitudes from -90\textdegree\ to 90\textdegree, azimuths from 0\textdegree\ to 360\textdegree\ and all integer distances for which that phase has a ray theoretical arrival predicted by ObsPy TauP in PREM \citep{Dziewonski1981}. These values therefore represent the full range of potential ellipticity corrections for these phases in PREM but true ranges might be different for  realistic source-receiver geometries in a specific application.

Ellipticity corrections are generally larger for S waves than P waves due to their lower velocities. For longer period body wave studies these corrections will be small relative to the wavelength of the signal, but for some studies these corrections are non-negligible compared to the magnitude of time anomalies that are interpreted. For high frequency studies ellipticity corrections should be routinely applied when interpreting absolute arrival times. This is especially true for more complex phases with long ray paths that have generally larger corrections.

For studies that use differential times, ellipticity corrections are likely to be less significant provided the phases have similar paths through the Earth. For SmKS differential times \citep[e.g.][]{Tanaka2007,wu2020}, the differential ellipticity corrections are on the order of hundreds to a couple of tenths of a second and are of minimal significance. However for phases with vastly different ray paths, for example P4KP - PcP differential time studies \citep[e.g.][]{Tanaka2010}, the differential ellipticity corrections are up to two seconds and non-negligible.

\subsection{Ellipticity corrections on Mars}

\begin{table*}
    \centering
    \caption{Spherical arrival times, ellipticity corrections and elliptical arrival times for seismic phases potentially detectable at the InSight Lander (4.50\textdegree N, 135.62\textdegree E) from a hypothetical event in  Cerberus Fossae (11.28\textdegree N, 166.37\textdegree E). \\}
    \begin{tabular}{|l|c|c|c|}
        \hline
	    Phase & \multicolumn{1}{|p{3cm}|}{\centering Spherical arrival time (s)} & Correction (s) & \multicolumn{1}{|p{3cm}|}{\centering Elliptical arrival time (s)} \\
        \hline
        P & 242.13 & 0.39 & 242.52\\
        \hline
        PP & 259.19 & 0.42 & 259.61\\
        \hline
        PcP & 401.87 & 0.71 & 402.57\\
        \hline
        S & 429.46 & 0.68 & 430.14\\
        \hline
        SS & 456.63 & 0.74 & 457.37\\
        \hline
        ScS & 740.85 & 1.30 & 742.15\\
        \hline
    \end{tabular}
    \label{CF_corr_table}
\end{table*}

We have implemented a velocity model of Mars \citep[model InSight\_KKS\_GP from][]{Knapmeyer2021,Khan2021,Stahler2021} in EllipticiPy in order to assess the effect of ellipticity of Mars on seismic waves. Mars is approximately twice as elliptical than Earth  at the surface \citep[][]{Bills1978} but its radius is approximately half that of Earth. Furthermore, the core fraction by radius is approximately similar to that of Earth \citep{Stahler2021}. The result of this combination is that ellipticity corrections for Mars have a similar magnitude to those for Earth but the travel times on Mars are smaller.

Figure \ref{Mars corrections} shows the minimum and maximum corrections for seismic phases that either have been detected on Mars or could potentially be detected on Mars: P, PP, S, SS \citep{Khan2021}, ScS \citep{Stahler2021} have been detected and PcP, P\textsubscript{diff} and S\textsubscript{diff} could potentially be detected in future work. These corrections are for a station fixed at the InSight lander location (4.50\textdegree N, 135.62\textdegree E) and with source depths between 0 and 50 km, as this is the range of source depths used in previous studies \citep[e.g.][]{Khan2021, Stahler2021}. These corrections cover a full range of back-azimuths and distances, however a large proportion of the events on Mars that have so far been located have been in the Cerberus Fossae region \citep[e.g.][]{Khan2021}. Ellipticity corrections and arrival times for an event located at the centre of the Cerberus Fossae region (11.28\textdegree N, 166.37\textdegree E) to the InSight lander are shown in Table \ref{CF_corr_table}. The magnitude of these corrections is negligible compared to the magnitude on the uncertainty of the phase picks by the Marsquake Service (MQS) which can be up to 60 seconds \citep[][]{Clinton2021}. Furthermore, due to the uncertainty of the source location, Martian studies to date have relied on differential times \cite[e.g.][]{Duran22,Drilleau21} for which ellipticity corrections are expected to be even less significant.

\section{Conclusions}

We have produced a Python package, EllipticiPy, that allows the easy calculation of ellipticity corrections that is designed to work alongside ObsPy TauP. This package is versatile and can calculate an ellipticity correction for any ray path that ObsPy TauP can return and removes the reliance on inaccurate and incomplete tables of ellipticity coefficients.

On Earth, ellipticity corrections are on the order of single seconds and should therefore be routinely applied when working with higher frequency seismic observations where trends on this order are to be interpreted. On Mars, ellipticity corrections are of approximately the same magnitude as on Earth, but are far smaller than the magnitude of the uncertainty on the phase picks given by the Marsquake Service. At present, ellipticity corrections on Mars are therefore negligible but this may not be the case in the future.

\section*{Data Availability}

Data analysis and production of figures was performed using Python, especially ObsPy \citep{Beyreuther2010}. The routines of \citet{Kennett1996} can be found at: \\ \url{https://github.com/GeoscienceAustralia/ellip-corr} in a convenient format within a Python wrapper. The coefficients of \citet{Kennett1996} can also be found in PDF format: \url{http://rses.anu.edu.au/seismology/AK135tables.pdf}.

For EllipticiPy, the source code, installation instructions, example usage and tables of coefficients can be found at this project's GitHub page: \url{https://github.com/StuartJRussell/EllipticiPy}.

\section*{Acknowledgments}

For the purpose of open access, the author has applied a Creative Commons Attribution (CC BY) licence to any Author Accepted Manuscript version arising from this submission. This project has received funding from the European Research Council (ERC) under the European Union’s Horizon 2020 research and innovation programme (grant agreement No. 804071 -ZoomDeep). JCEI acknowledges support from UKSA grant ST/W002515/1. The synthetic modelling in this work was performed using resources provided by the Cambridge Service for Data Driven Discovery (CSD3) operated by the University of Cambridge Research Computing Service (\url{www.csd3.cam.ac.uk}), provided by Dell EMC and Intel using Tier-2 funding from the Engineering and Physical Sciences Research Council (capital grant EP/T022159/1), and DiRAC funding from the Science and Technology Facilities Council (\url{www.dirac.ac.uk}).

The authors would like to thank past and present members of the University of Cambridge Global Seismology research group, especially Carl Martin, George Pindar, Alistair Boyce and Florian Millet for helpful and insightful scientific discussions. The authors would also like to thank Brian Kennett and Olafur Gudmundsson for their personal communications regarding their previous works. The authors would also like to thank the editor, Carl Tape, as well as Istvan Bondar and one anonymous reviewer for their constructive comments and helping to improve the original manuscript.

\bibliography{bibliography}

\end{document}


\maketitle

This document includes:
\begin{description}
    \item \hspace{0.5cm} \ref{correction_derivation}. Derivation of ellipticity corrections for seismic phases.
    
    \item \hspace{0.5cm} \ref{ellipticity_derivation}. Derivation of the ellipticity of figure with radius for a rotating body in hydrostatic equilibrium.
    
    \item \hspace{0.5cm} \ref{min_max}. Tables of minimum and maximum ellipticity corrections for seismic phases on Earth.
    
    \item \hspace{0.5cm} \ref{coeff tables}.
    Information regarding tables of ellipticity correction coefficients.
    
\end{description}

\pagebreak
\section{Derivation of Ellipticity Corrections} \label{correction_derivation}

The following derivation obtains the result presented in the theory section of the main publication. The derivation is essentially that of \citet{Dziewonski1976} coupled with the reformulation of \citet{Doornbos1988}.

In spherical coordinates, a point can be defined at a radius, $r$, co-latitude, $\vartheta$, and longitude, $\phi$. A surface of constant velocity can then be defined as

\begin{equation} \label{surface_def}
    R(\vartheta,\phi) = r + \delta r (\vartheta,\phi)
\end{equation}

\noindent where $r$ is the spherically averaged value of $R(\vartheta,\phi)$

\begin{equation} \label{spherical_average}
    r = \frac{1}{4\pi} \int_{S} R(\vartheta,\phi) \,\d\Omega.
\end{equation}

\noindent The velocity, $v$, on this surface is constant

\begin{equation} \label{constant_velocity}
    v(R(\vartheta,\phi)) = v(r + \delta r (\vartheta,\phi)) = \text{constant}
\end{equation}

\noindent and by a Taylor expansion

\begin{equation} \label{taylor}
    v(r + \delta r (\vartheta,\phi)) = v(r) + \delta r \, v ^\prime (r) + \mathcal{O}(\delta r^2) = \text{constant}.
\end{equation}

\noindent The velocity at $r$ can be defined as the velocity at the surface, $R$, plus a small perturbation, $\delta v$

\begin{equation} \label{vel_pertubation}
    v(r, \vartheta, \phi) = v(R(\vartheta,\phi)) + \delta v(r, \vartheta, \phi).
\end{equation}

\noindent By substituting \eqref{taylor} into \eqref{vel_pertubation} it is possible to define $\delta v$

\begin{equation} \label{dvel}
    \delta v(r, \vartheta, \phi) = -v^\prime{}(r) \delta r (\vartheta, \phi).
\end{equation}

\noindent The relative perturbation of a surface is defined by

\begin{equation} \label{sigma}
    \varsigma(r, \vartheta, \phi) \equiv r^{-1}\delta r (\vartheta, \phi),
\end{equation}

\noindent and can be related to a planet's ellipticity of figure $\epsilon$ by

\begin{equation} \label{ellipticity}
    \varsigma(r, \vartheta, \phi) = \epsilon(r)\left(1/3 - \cos^{2}\vartheta\right) = -\frac{2}{3}\epsilon(r) P_{2}^{0}(\cos\vartheta).
\end{equation}

\noindent where $P_l^m(x)$ denotes associated Legendre polynomials of degree $l$ and order $m$. For seismic waves propagating within an elliptical Earth it is necessary to analyse the effect of $\delta v$ on the travel time of a ray. The angular slowness of a ray, $p$, is constant. For a point on a ray with epicentral coordinates, $(r,\theta)$,

\begin{equation} \label{slowness}
    dt/d\theta = p = r v^{-1} \sin i = \eta \sin i
\end{equation}

\noindent where $i$ is the incident angle (angle from vertical) of the ray, and $\eta$ is the ray slowness defined by

\begin{equation}
    \eta \equiv \frac{r}{v}.
\end{equation}

\noindent The vertical slowness, which varies along the ray, is defined as

\begin{equation} \label{vertslow}
    q \equiv \eta \cos i \equiv (\eta^{2} - p^{2})^{1/2}.
\end{equation}

If a point moves a small distance, $\delta d$, along the ray path then the radial distance moved is $\delta r = \delta d/\cos i$. Therefore by using \eqref{slowness} and  \eqref{vertslow} to relate $p$, $\eta$ and $q$, the time taken, $\delta t$, is

\begin{equation} \label{dt_on_ray}
    \delta t = v^{-1}\delta d = v^{-1}  \cos i\, \delta r =  \eta r^{-1} \cos i \,  \delta r=  \,q r^{-1} \delta r.
\end{equation}

\noindent Let the following notation represent the jump in value of a function, $z(r)$, across a surface, $r$:

\begin{equation} \label{limit}
    \left[z(r)\right]_-^+ = \lim_{s \to 0^+} \left[z(r+s) - z(r-s)\right].
\end{equation}

\noindent It is then possible to express the effect of moving a discontinuity in $v$ by a small amount $\delta r$. This is done by taking the difference of \eqref{dt_on_ray} evaluated for a ray travelling between $r + \delta r$ and $r$ in the case of both the upper and lower layers (propagating in the other layer between $r + \delta r$ and $r$ is the simplest way to imagine the effect of moving the discontinuity between the two layers). For transmission this is 

\begin{equation} \label{transmission}
    \delta t = -r^{-1} \delta r\left[q\right]_-^+ = -\varsigma\left[q\right]_-^+
\end{equation}

\noindent and for reflection is

\begin{equation} \label{reflection}
    \delta t = \mp r^{-1} \delta r\left(q_0 + q_1\right) = \mp \varsigma \left(q_0 + q_1\right)
\end{equation}

\noindent where the preceding minus/plus refers to top-side/bottom-side reflections, respectively, and $\eta_{0}$ and $\eta_{1}$ represent the value of $\eta$ before and after interaction with the boundary. In the case where there is no conversion of wave type due to interaction with the boundary, equation \eqref{reflection} simplifies to

\begin{equation} \label{reflection_simple}
    \delta t = \mp 2 \varsigma q.
\end{equation}

\noindent The source depth must also be deformed to account for the effects of ellipticity, and this is done using the above equations, treating the source depth as a discontinuity where the wave does not propagate in the upper medium (or lower medium in the case of phases with an upwards take-off angle, such as those preceded by a p or s segment).

The following equation is a result from \citet{Julian1968} (their equation 4) for the effect on travel time, $\delta t$,  of a small perturbation in velocity, $\delta v$, on a ray, by integrating between the bottoming point of the ray, $r_{p}$, and the radius of the Earth, $a$

\begin{equation} \label{Julian}
    \delta t = - \int_\text{ray path} v^{-2} \delta v \, \d d = -2\int_{r_{p}} ^{a} \eta v^{-2} q^{-1} \delta v \,\d r.
\end{equation}

\noindent The above can be generalised for a source at $r_{0}$ and a receiver at $r_{1}$ (for a receiver at the surface $r_{1} = a$)

\begin{equation} \label{general_Julian}
    \delta t = - \sum_{n=0}^{1} {\int_{r_{p}}^{r_{n}} \eta v^{-2} q^{-1} \delta v \,\d r},
\end{equation}

\noindent which on substituting \eqref{dvel} and \eqref{sigma} becomes 

\begin{equation}
    \label{radial_integral}
    \delta t = \sum_{n=0}^{1} {\int_{r_{p}}^{r_{n}} \eta v^{-2} q^{-1} \varsigma v^{\prime} r \,\d r}.
\end{equation}

\noindent \citet{Dziewonski1976} transform equation \eqref{radial_integral} into an integral over distance

\begin{equation} \label{sigma_integral_simple}
    \delta t = p^{-1} \int_{0} ^{\Delta} \eta^{3}  \varsigma v^{\prime} \,\d\theta,
\end{equation}

\noindent where $\Delta$ is the total epicentral distance. However, for ray paths that propagate close to the centre of the planet, equation \eqref{sigma_integral_simple} can be problematic due to a discontinuity in $\theta$ that occurs at the centre of the planet. We instead rewrite \eqref{radial_integral} using the vertical slowness, noting the derivatives,

\begin{gather}
    \frac{\d q}{\d \eta} = \frac{\eta}{q}, \\
    \frac{\d \eta}{\d r} = \frac{1}{v} - \frac{r}{v^{2}} v^{\prime},
\end{gather}

\noindent from which it follows that \eqref{radial_integral} can be written as

\begin{equation} \label{slowness_integral}
    \delta t = \sum_{i} \left\{ {\int_{q_{0}}^{q_{1}}  \frac{\eta v^{\prime}}{1 - \eta v^{\prime}} \ \varsigma \ \d q} \right\}_{i}
\end{equation}

\noindent which is an integration over vertical slowness for each continuous section, $i$, of a ray path where $q_{0}$ and $q_{1}$ are the values of $q$ at the start and end of each ray segment, with $q_1>q_0$. \eqref{slowness_integral}. It is convenient to introduce the parameter, $\xi$, from \citet{Bullen1963}:

\begin{equation} \label{xi}
    \xi = \frac{\eta}{r} \left( \frac{\d \eta}{\d r} \right)^{-1} = \frac{\d \ln r}{\d \ln \eta}.
\end{equation}

\noindent Equation \eqref{slowness_integral} therefore becomes

\begin{equation} \label{slowness_integral_simplified}
    \delta t = \sum_{i} \left\{ {\int_{q_{0}}^{q_{1}}  \left( \xi - 1 \right) \varsigma \ \d q} \right\}_{i}
\end{equation}

\noindent which is similar to the formulation of \citet{Doornbos1988}. The parameter $\xi$ is constant when using the Mohorovicic or Bullen law $v(r) = A r^B$ for interpolation within a velocity layer \citep{Buland1983}. 

Using the cosine rule for the surface of a sphere, the co-latitude, $\vartheta$, at any distance, $\theta$, on a ray path with a given azimuth, $\zeta$, can be determined

\begin{equation} \label{ray_colat}
    \cos \vartheta = \cos \vartheta_{0} \cos \theta + \sin \vartheta_{0} \sin \theta \cos \zeta
\end{equation}

\noindent Equation \eqref{ray_colat} is in the correct form to apply the Addition Theorem for Legendre Functions \citep{Arfken1985}

\begin{equation} \label{legendre_addition}
    P_{2}^{0}(\cos\vartheta) = \sum_{m=0}^{2} P_{2, m}(\cos \vartheta_{0}) P_{2,m} (\cos \theta) \cos m \zeta,
\end{equation}

\noindent where

\begin{equation} \label{weighted_legendre}
    P_{l, m}(\cos \theta) = \left[(2 - \delta_{m, 0}) \frac{(l-m)!}{(l+m)!}\right]^{1/2} P_{l}^{m}(\cos \theta).
\end{equation}

\noindent $P_{l, m}(x)$ are the Schmidt semi-normalised associated Legendre polynomials, where the normalising weights are such that the Addition Theorem is simplified. 

\noindent Using equations \eqref{ellipticity} and \eqref{legendre_addition} it follows that

\begin{equation} \label{sigma_legendre_addition}
    \varsigma = -\frac{2}{3} \epsilon \sum_{m=0}^{2} P_{2, m} (\cos\vartheta_{0}) P_{2, m} (\cos \theta) \cos m \zeta.
\end{equation}

\noindent Substituting equation \eqref{sigma_legendre_addition} into equation \eqref{slowness_integral} gives

\begin{equation} \label{full_legendre_addition}
    \delta t = - \frac{2}{3} \sum_{m=0}^{2} P_{2, m}(\cos \vartheta_{0}) \cos m \zeta {\int_{q_{0}}^{q_{1}} \left( \xi - 1 \right) \epsilon P_{2,m}(\cos \theta) \ \d q}
\end{equation}

\noindent It is possible to include the effects of phase and distance in calculable coefficients, $\sigma_{m}$. Rewriting equation \eqref{full_legendre_addition} as

\begin{equation} \label{tau_legendre_addition}
    \delta t = \sum_{m=0}^{2} \sigma_{m} P_{2, m}(\cos \vartheta_{0}) \cos m \zeta
\end{equation}

\noindent where the ellipticity correction coefficients, $\sigma_{m}$, also include the effects of displacing discontinuities in $v$ as per equations \eqref{transmission} and \eqref{reflection}

\begin{equation} \label{tau}
            \sigma_{m} = \sum_{i} \left\{ {\int_{q_{0}}^{q_{1}} \left( \xi - 1 \right) \epsilon \lambda_m (\theta) \ \d q} \right\}_{i} -  \sum_{j} \left\{ \epsilon \lambda _{m} (\theta) \left[q\right]_-^+ \right\}_{j} \mp \sum_{k} \left\{ \epsilon \lambda _{m} (\theta) q \right\}_{k}
\end{equation}

\noindent where $i$ represents a sum over continuous regions of the ray path and $j$ and $k$ represent sums over all transmitting and reflecting boundaries, respectively, and

\begin{equation} \label{lambda}
    \lambda_{m}(\theta) = - \frac{2}{3} P_{2, m}(\cos \theta).
\end{equation}

Equation \eqref{tau} is for a phase that remains one wave type along its entire path; it is important, that in the case where phase conversions at boundaries take place, that the equations for transmission and reflection through boundaries are applied correctly. For diffracted phases, the change in arc length due to ellipticity is second order in $\epsilon$ \citep{Kennett1996} and can be ignored. For diffracted phases it is therefore only necessary to apply equation \eqref{tau} to the up-going and down-going parts of the ray path.

\pagebreak
\section{Ellipticity of Surfaces of Constant Density} \label{ellipticity_derivation}

For any planet with ellipticity due to rotation, the relative perturbation of a surface, $\delta r$, from the reference radius of the surface, $r$, can be related to co-latitude, $\vartheta$, by equation \eqref{ellipticity}. The ellipticity of figure for a surface of equal density, $\epsilon$, varies with depth and is dependent on the density distribution within a planet and its angular velocity. It is necessary to calculate $\epsilon$ before ellipticity corrections for seismic waves can be calculated for that velocity model. The following is only a brief overview of the method to calculate $\epsilon$ and a full derivation can be found in \citet{Bullen1975}. An application to simple Earth models can be found in \citet{Bullen1973}.

Given a density distribution of a body with radius, $\rho(r)$, it is possible to calculate the mass, $M(r)$, within a surface of radius $r$ for a spherical body by an integral of volume:

\begin{equation} \label{mass}
    M(r) = \int \rho \,\d V = 4 \pi \int_{0}^{r} \rho r^{2} \,\d r.
\end{equation}

\noindent The moment of inertia, $I(r)$, of a body with radial density distribution, $\rho(r)$, can be calculated by treating the body as the sum of thin spherical shells:

\begin{equation} \label{MOI}
    I(r) = \frac{2}{3} \pi \int \rho r^{2} \,\d V = \frac{8}{3} \pi \int_{0}^{r} \rho r^{4} \,\d r.
\end{equation}

\noindent The mean moment of inertia coefficient, $y(r)$, for the mass within each surface of constant density is then calculable

\begin{equation} \label{y}
    y(r) = \frac{I(r)}{M(r) r^{2}}.
\end{equation}

\noindent Using the Darwin-Radau approximation, which assumes hydrostatic equilibrium, the Radau parameter, $\upeta$, and $y$ are related by 

\begin{equation} \label{radau}
    \frac{2}{5} (1 + \upeta)^{1/2} = 1 - \frac{3}{2} y.
\end{equation}

\noindent Radau's parameter is related to ellipticity by a differential equation:

\begin{equation} \label{diff}
    \upeta = \frac{r}{\epsilon} \frac{\d\epsilon}{\d r}
\end{equation}

\noindent which can be solved by separation of variables

\begin{equation} \label{solution}
    \epsilon =  c \exp \left({\int \frac{\upeta}{r}\, \d r}\right)
\end{equation}

\noindent where $c$ is a constant of integration. Using the Darwin-Radau approximation it is possible to estimate $\epsilon$ at the surface. Let the subscript $a$ represent variables at the surface.

\begin{equation} \label{eta_surface}
    \upeta_{a} = \frac{5h}{2 \epsilon_{a}} - 2
\end{equation}

\noindent where $h$ is a convenient geodynamical parameter representing the ratio of rotational and gravitational forces at the equator on the surface of a body,

\begin{equation} \label{h}
    h = \frac{a^{3} \Omega^{2}}{GM},
\end{equation}

\noindent where $a$ is the radius of the body, $\Omega$ is the angular velocity of the body, $M$ is the mass of the body (calculated using equation \eqref{mass}) and $G$ is the universal gravitational constant. Having calculated $\epsilon_{a}$ it is possible to calculate the constant of integration, $c$ in equation \eqref{solution}.

This formulation allows the easy calculation of $\epsilon$ for any surface of equal density within a rotating body with a given density distribution where Darwin-Radau's equation holds. This assumes that the body is in hydrostatic equilibrium and is an ellipsoid of revolution. This assumption does not hold for all bodies, notably the gas giants of our own solar system, however the practical difficulties of doing seismology on such bodies means that the need for ellipticity corrections in such a scenario is unlikely. For the vast majority of bodies where seismology is likely to be performed, this formulation is sufficient.

\pagebreak
\section{Minimum and Maximum Ellipticity Corrections on Earth} \label{min_max}

Table \ref{Earth_min_max_table} shows the minimum and maximum ellipticity corrections for various seismic phases in PREM \citep{Dziewonski1981}. These correspond to those shown in Figure 3 in the main publication and were calculated by a systematic search with source depths from 0 km to 700 km, source latitudes from -90\textdegree\ to 90\textdegree\, azimuths from 0\textdegree\ to 360\textdegree\ and all integer distances for which that phase has a ray theoretical arrival predicted by ObsPy TauP \citep{Crotwell1999, Beyreuther2010} in PREM.

\begin{table}[H]
	\centering
    \caption{Minimum and maximum ellipticity corrections for commonly used seismic phases in PREM. Corrections for depth phases can be calculated but are not shown here. \\}
	\begin{tabular}{|l|c|c|c|}
		\hline
		Phase & Distance range & Minimum correction (s) & Maximum Correction (s) \\ 
		\hline
		p (up-going) & 0\textdegree\ - 12\textdegree & -0.44 & 0.24 \\ 
		\hline
		P & 0\textdegree\ - 98\textdegree & -1.01 & 0.97 \\ 
		\hline
		PP & 0\textdegree\ - 196\textdegree & -1.75 & 1.94 \\ 
		\hline
		PcP & 0\textdegree\ - 98\textdegree & -1.49 & 0.97 \\ 
		\hline
		Pdiff & 96\textdegree\ - 158\textdegree & -1.86 & 0.97 \\ 
		\hline
		PKPab & 143\textdegree\ - 177\textdegree & -2.19 & 1.39 \\ 
		\hline
		PKPbc & 143\textdegree\ - 152\textdegree & -2.24 & 1.32 \\ 
		\hline
		PKPdf & 116\textdegree\ - 180\textdegree & -2.68 & 1.34 \\ 
		\hline
		PKKPab & 235\textdegree\ - 257\textdegree & -1.37 & 1.82 \\ 
		\hline
		PKKPbc & 235\textdegree\ - 282\textdegree & -2.19 & 1.89 \\ 
		\hline
		PKKPdf & 210\textdegree\ - 360\textdegree & -3.88 & 1.94 \\ 
		\hline
		P'P'ab & 289\textdegree\ - 355\textdegree & -4.36 & 2.79 \\ 
		\hline
		P'P'bc & 289\textdegree\ - 305\textdegree & -4.45 & 2.64 \\ 
		\hline
		PKiKP & 0\textdegree\ - 152\textdegree & -2.33 & 1.32 \\ 
		\hline
		s (up-going) & 0\textdegree\ - 11\textdegree & -0.78 & 0.43 \\ 
		\hline
		S & 0\textdegree\ - 102\textdegree & -1.83 & 1.81 \\ 
		\hline
		SS & 0\textdegree\ - 205\textdegree & -3.18 & 3.62 \\ 
		\hline
		ScS & 0\textdegree\ - 102\textdegree & -2.72 & 1.81 \\ 
		\hline
		Sdiff & 100\textdegree\ - 162\textdegree & -3.53 & 1.81 \\ 
		\hline
		SKSac & 60\textdegree\ - 141\textdegree & -3.19 & 1.93 \\ 
		\hline
		SKSdf & 106\textdegree\ - 180\textdegree & -3.92 & 1.96 \\ 
		\hline
		SKKSac & 60\textdegree\ - 271\textdegree & -3.34 & 2.50 \\ 
		\hline
		SKKSdf & 201\textdegree\ - 360\textdegree & -5.11 & 2.56 \\ 
		\hline
		S'S'ac & 122\textdegree\ - 283\textdegree & -5.59 & 3.87 \\ 
		\hline
		SKiKS & 0\textdegree\ - 141\textdegree & -3.56 & 1.93 \\ 
		\hline
		PS & 18\textdegree\ - 141\textdegree & -2.43 & 2.35 \\ 
		\hline
		PcS & 0\textdegree\ - 62\textdegree & -2.11 & 1.18 \\ 
		\hline
		PKSab & 129\textdegree\ - 142\textdegree & -2.39 & 1.61 \\ 
		\hline
		PKSbc & 129\textdegree\ - 147\textdegree & -2.78 & 1.63 \\ 
		\hline
		PKSdf & 110\textdegree\ - 180\textdegree & -3.30 & 1.65 \\ 
		\hline
		PKKSab & 213\textdegree\ - 222\textdegree & -2.28 & 2.04 \\ 
		\hline
		PKKSbc & 214\textdegree\ - 276\textdegree & -2.32 & 2.19 \\ 
		\hline
		PKKSdf & 205\textdegree\ - 360\textdegree & -4.49 & 2.25 \\ 
		\hline
		SP & 12\textdegree\ - 141\textdegree & -2.46 & 2.35 \\ 
		\hline
		ScP & 0\textdegree\ - 62\textdegree & -2.11 & 1.18 \\ 
		\hline
		SKPab & 130\textdegree\ - 142\textdegree & -2.33 & 1.61 \\ 
		\hline
		SKPbc & 130\textdegree\ - 147\textdegree & -2.72 & 1.63 \\ 
		\hline
		SKPdf & 111\textdegree\ - 180\textdegree & -3.30 & 1.65 \\ 
		\hline
		SKKPab & 215\textdegree\ - 222\textdegree & -2.22 & 2.04 \\ 
		\hline
		SKKPbc & 215\textdegree\ - 276\textdegree & -2.34 & 2.19 \\ 
		\hline
		SKKPdf & 206\textdegree\ - 360\textdegree & -4.49 & 2.25 \\ 
		\hline
	\end{tabular}
	\label{Earth_min_max_table}
\end{table}

\pagebreak
\section{Tables of ellipticity correction coefficients} \label{coeff tables}

While direct calculation of ellipticity corrections for a given ray path is expected to  be the most accurate and convenient method of calculating ellipticity corrections, in applications where large numbers of sources and receivers are used it may be substantially more efficient to interpolate pre-calculated tables of coefficients. We have produced tables of coefficients similar to those of \cite{Kennett1996} as available in their \texttt{ELCOR.dat} file.

The original \texttt{ELCOR.dat} contains coefficients sampled at 5 degree distance intervals and irregular depth intervals (0 km, 100 km, 200 km, 300 km, 500 km, 700 km) for a range of seismic phases in the ak135 velocity model \citep{Kennett1995}. The phase names, order and dimensions of this file are hardcoded in the original \texttt{ellip} package and therefore replacement tables must contain the same phases in the same order with the same number of entries per phase in order to be compatible.

One complication of the original \texttt{ELCOR.dat} is that it contains coefficients even where there is not a ray-theoretical arrival and this is highlighted in the \texttt{ellipcor.help} file of the \texttt{ellip} package: \textit{`The presence of a value in the tables does not imply a physical arrival at all distance, depth combinations. Where necessary extrapolation has been used to ensure satisfactory results.'} Exactly what is meant by `necessary extrapolation' is not apparent.

Using EllipticiPy, we have produced a replacement \texttt{ELCOR.dat} file, called \texttt{ELCOR\_replacement.dat}, that is designed to act as a direct replacement. These coefficients are for the ak135 model and are sampled at the same distance and depth points as the original file. The various assumptions and procedures used to produce this file are:

\begin{itemize}
    \item For points where a phase name has several possible ray paths (e.g. for upper mantle P wave triplications) coefficients for the first arriving ray path are provided.
    
    \item Branches of core phases are separated as distinct phases, except for P$'$P$'$ and S$'$S$'$ which are explained below.
    
    \item For points where we do not expect a ray-theoretical arrival but there is reason to believe that the original values from \texttt{ELCOR.dat} are correct (phases below 180 degrees and good agreement where there are arrivals) we have used the original values.
    
    \item For points where we do not expect a ray-theoretical arrival but there is reason to believe that the original values from \texttt{ELCOR.dat} are in doubt, we have linearly extrapolated from the closest calculable values.
    
    \item In the original \texttt{ELCOR.dat}, tables for P$'$P$'$ and S$'$S$'$ exceed the expected distance ranges for those phases that bottom in the outer core, and in parts of the given range correspond to the df (inner core) branches of the phase. In order to keep \texttt{ELCOR\_replacement.dat} compatible, this file must have the same number of entries for each phase as the original \texttt{ELCOR.dat} file. Therefore for P$'$P$'$ the coefficients given are for P$'$P$'$df and for S$'$S$'$ they are for S$'$S$'$ac at the lower end of the distance range and for S$'$S$'$df at the upper end of the range. The coefficients for P$'$P$'$bc and P$'$P$'$df are continuous, as are those for S$'$S$'$ac and S$'$S$'$df. The only branch not represented by these phase entries is P$'$P$'$ab.
    
    \item PnS as we understand it (a P wave that descends to and diffracts along the Moho before continuing downwards as an S wave) does not occur totally within the distance range given in the original \texttt{ELCOR.dat}, nor do the coefficients given match those calculated for this phase. As we do not know what ray path these coefficients are for and cannot reproduce them at the exact distances, we omit this phase in \texttt{ELCOR\_replacement.dat}. 
\end{itemize}

One major disadvantage of using pre-calculated tables is that the exact ray path used to calculate the coefficients is not known to the user. Therefore if in doubt, we encourage all users to use EllipticiPy directly as ObsPy allows easy visualisation of ray paths to confirm that the coefficients are for the intended ray path. EllipticiPy also contains a function which produces tabulated coefficients for a given phase and source depth. Coefficients are produced for the full range of ray parameters for which the phase exists, using the same discretisation in slowness as used for the velocity model by ObsPy TauP. Such coefficients can be pre-calculated and then used in a standard interpolation routine. 

\texttt{ELCOR\_replacement.dat} is available at this project's GitHub repository, \url{https://github.com/StuartJRussell/EllipticiPy}, along with the main python package. 

\pagebreak
\bibliography{bibliography}